\documentclass[pdftex]{article}

\usepackage{spconf,amsmath,graphicx}
\usepackage[matnotit]{echodefs}
\usepackage{booktabs}
\usepackage{microtype}

\usepackage{mathabx}

\newcommand{\proj}{\mathsf{P}}
\newcommand{\affest}{\mathsf{R}}

\newcommand{\kurt}{\mathsf{kurt}}
\newcommand{\var}{\mathsf{\mathbb{V}}}

\newcommand{\normtv}[1]{\norm{#1}_{\mathsf{TV}}}
\newcommand{\corr}{\mathsf{corr}}

\graphicspath{ {./graphics/} }

\title{Inverse Problems with Invariant Multiscale Statistics}
\name{%
Ivan Dokmani\'c$^1$, Joan Bruna$^2$, St\'ephane Mallat$^3$, and Maarten de Hoop$^4$
\thanks{This work was supported by the ERC grant InvariantClass 320959. ID was funded by LABEX WIFI (Laboratory of Excellence within the French Program ``Investments for the Future'') under references ANR-10-LABX-24 and ANR-10-IDEX-0001-02 PSL* and by Agence Nationale de la Recherche under reference ANR-13-JS09-0001-01. We would like to thank Laurent Daudet for valuable discussions in the early stages of this research.}}
\address{%
\begin{tabular}{ll}
$^1$Institut Langevin, CNRS, ESPCI, PSL, Paris, France   &  $^2$Courant Institute, NYU, New York, USA \\
$^3$Ecole Normale Superieure, PSL, Paris, France         &  $^4$CAAM, Rice University, Houston, USA
\end{tabular}
\vspace{-2mm}
}

\begin{document}
\ninept
\maketitle
\begin{abstract}
    We propose a new approach to linear ill-posed inverse problems. Our algorithm alternates between enforcing two constraints: the measurements and the statistical correlation structure in some transformed space. We use a non-linear multiscale scattering transform which discards the phase and thus exposes strong spectral correlations otherwise hidden beneath the phase fluctuations. As a result, both constraints may be put into effect by linear projections in their respective spaces. We apply the algorithm to super-resolution and tomography and show that it outperforms ad hoc convex regularizers and stably recovers the missing spectrum.
\end{abstract}
\begin{keywords}%
Inverse problems, tomography, super-resolution, scattering transform, regularization.
\end{keywords}
\section{Introduction}
\label{sec:intro}

In this paper we propose a new way to solve ill-posed linear inverse problems. We consider the usual problem statement: estimate $x \in {\cal X}$ from noisy measurements $y$ by a singular operator $\Gamma$:

\begin{equation}
    \label{eq:measurements}
    y = \Gamma x + b,
\end{equation}
with $b$ being the measurement noise. The estimator $\wh{x} = \wh{x}(y)$ of $x$ is a function of the measurements $y$.

A standard approach to deal with ill-posedness is to cook up a regularizer $h(u)$ which promotes signals within the desired model and solve
\[
    \min_{u \in {\cal X}} \ d(y, \Gamma u) + \lambda h(u),
\]
where $d(y, \Gamma u)$ is the data fidelity term \cite{Engl:2000ul,Daubechies:2004em}. Both $d$ and $h$ are typically convex which may be undesirable in severely ill-posed problems (cf. Section \ref{sub:comparison_with_l1}).

\begin{figure}
    \label{fig:known_spectrum}
    \centering
    \includegraphics[width=\linewidth]{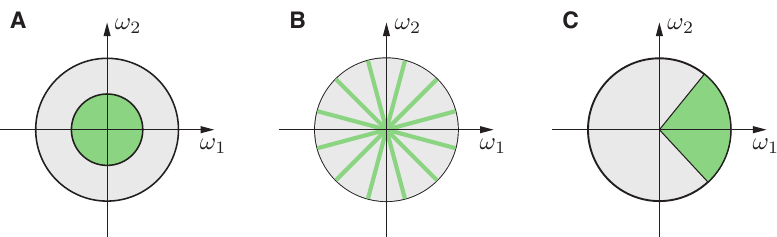}
    \caption{Known part of the Fourier transform with super-resolution measurements (A); Radon transform uniformly subsampled in angle (B); Radon transform with directions restricted to a cone (C).}
\end{figure}

We propose a different route. Measurements $y$ reveal certain information about the signal $x$ that can often be interpreted spectrally: a part of the spectrum is known from $y$, and solving the inverse problem means reconstructing the unknown part from the known part. This is illustrated in Fig. \ref{fig:known_spectrum} for the two inverse problems we study---super-resolution and tomography. Any signal can be made to satisfy the measurements by projecting it in an appropriate space ${\cal A}$ of signals consistent with measurements. For example, in the case of super-resolution, this projection replaces the low frequencies by the measured ones.

For some $z$, $\proj_{\cal A}(z)$ may be a bad estimate of $x$ in the sense that the joint statistics of the known and unknown spectral regions will be wrong. We show that these joint statistics are captured by the \emph{scattering coefficients} \cite{Bruna:2013hy,Mallat:2012by}---local averages of moduli of complex wavelet coefficients which encode interactions across scales. Scattering has been used earlier to build probabilistic models for super-resolution \cite{Bruna:2015un}.

To see why we work with amplitudes and discard the phases, consider estimating a complex-valued random variable $X = \abs{X} \e^{j \phi_X}$ from $Y = \abs{Y} \e^{j \phi_Y}$. Assume that phases and magnitudes are independent, $\phi_X$ is distributed uniformly on a circle, and phases are perfectly correlated: $\phi_Y = -\phi_X$. One may expect that this correlation is beneficial. However, the quality of linear estimation depends on achieving significant linear correlations between coordinates of the transform, as measured by the correlation coefficients \cite{Pearson:1895bj}:\footnote{Even in the Gaussian case, absence of correlation in the sense of \eqref{eq:correlation} limits our estimation potential to that of restoring the mean.}
\begin{equation}
    \label{eq:correlation}
    \corr(X_i, X_j) = \frac{\E[(X_i - \E[X_i])(X_j - \E[X_j]))}{\sqrt{\var(X_i) \var(X_j)}}.
\end{equation}
A quick computation shows that $\E[XY^*] = 0$, so that correlation is zero and at best we can restore the mean ($0$ in this case). In general, phases inevitably reduce linear correlations: an instance of a phenomenon known in statistics as regression to the mean.

By eliminating the phase, a scattering transform yields strongly correlated coefficients which can thus be estimated \emph{linearly} from the scattering coefficients of the measured signal. 
This estimator defines another projection in the space of square integrable random variables. Signals obtained by this second projection will generally not belong to $\cal A$. We then propose a natural idea: iterate the two projections---a linear one onto $\cal{A}$ and a non-linear one that adjusts the transform statistics---until we obtain a signal in $\cal A$ with the right statistics.

We demonstrate the method in two inverse problems: super-resolution and tomography, showing that our approach stably estimates the missing spectral information. The paper is organized as follows: Section 2 reviews scattering transforms, Section 3 presents our main approach and Section 4 gives the algorithmic details. Finally, Section 5 presents numerical experiments.


\section{Scattering Transform}
\label{sec:scattering}

A scattering representation is a vector of signal descriptors computed as moduli of iterated wavelet coefficients, which are made locally invariant to translations. By eliminating the phase, it creates a set of highly correlated coefficients so that the missing spectrum can be linearly estimated from known coefficients at different scales.

Let $x(u)$ for $u \in \Z^d$ be a stationary process, with $\E [|x|^2] < \infty$. A wavelet transform analyzes multiscale variability of $x$ through convolutions with dilated wavelet band-pass filters. For image processing, the wavelets $\psi_k$ may be constructed by rotating a single
wavelet $\psi$ throughs an angle $k \pi/K$: $\psi_k (u) = \psi(r_k u)$.
They are dilated diadically by factors of $2^j$, 
$\psi_{j,k} (u) = 2^{-j n} \psi_k (2^{-j} u)$, up to the maximum
scale $2^J$. The resulting wavelet transform of $x$ is
\begin{equation}
\label{wave1}
{\cal W}_J x = \{ x \star \phi_J(u) \, , \, x \star \psi_{j,k}
(u)\}_{j \leq J , 1 \leq k \leq K}~,
\end{equation}
where 
$\phi_J (u) = 2^{-dJ} \phi (2^{-J} u)$ is a scaled averaging filter, with
$\int \phi(u)\,du = 1$.

If $x$ is stationary then $x \star \psi_{j,k} (u)$ is also stationary. We consider complex-valued wavelets $\psi$ whose Fourier transform has a support concentrated on one-half of the Fourier domain---for any $\omega \in \R^d$ either $\hat \psi(\omega) \approx 0$ or $\hat \psi(-\omega) \approx 0$. We define $K$ wavelets $\psi_k (u)$, which are regular functions with a fast decay and a zero average $\int \psi_k (u)\, du = 0$. One can verify that the wavelet transform is invertible with a stable inverse if the wavelet Fourier transform satisfies the following Littlewood-Paley condition at all frequencies $\omega$:

\begin{equation}
    \label{Littepsdofnsd}
    1 - \epsilon \leq |\phi_J (\omega)|^2 + \frac 1 2 \,\sum_{j,k}  (|\hat \psi_{j,k} (\omega)|^2 + |\hat \psi_{j,k} (-\omega)|^2 ) \leq 1~.
\end{equation}
First-order scattering coefficients are obtained by averaging the envelope of the complex wavelet coefficients:

\begin{equation}
\label{scat2}
\Phi x(u,j,k) = |x \star \psi_{j,k}| \star \phi_J (u) ~.
\end{equation}
Because of the averaging, these coefficients are locally invariant to translations at a scale $2^J$. However, the averaging erases information, and to recover it we must recover the variability of the wavelet transform envelope $|x \star \psi_{j,k}(u)|$ for each $j,k$. 

Second-order scattering coefficients provide information about this variability by filtering the envelope with a new set of wavelets $\psi_{j',k'}$ and averaging the amplitude of the resulting coefficients~with~$\phi_J$:
\[
\Phi x(u,j,k,j',k') = ||x \star \psi_{j,k}| \star \psi_{j',k'}| \star \phi_J (u)~.
\]
They measure the average multiscale time variations of $|x \star \psi_{j,k}(u)|$. Note that in general, the scattering transform is not invertible.

If $x(u)$ is a stationary process then $|x \star \psi_{j,k}| \star \phi_J(u)$ and $||x \star \psi_{j,k}| \star \psi_{j',k'}| \star \phi_J (u)$ are also stationary, but also slowly varying in space. They are strongly correlated with correlation coefficients which depend upon the properties of $x$. In the following we explain how learning these correlation coefficients enables us to estimate the missing spectral information.


\section{Our Approach in a Nutshell} 

As mentioned in the introduction, we work with two types of information about $x$: measurements $y$ and the correlation structure of our representation $\Phi x$. Satisfying the measurement constraints is achieved with a linear projection in the 
original signal domain, whereas adjusting the correlation structure is achieved with a linear operator in the transformed domain $\Phi$. We assume knowing the first and second moments of $(\Phi x, \Phi \Gamma^+ y)$.
%
%

In general, the two criteria will not have the same minimizers, therefore the estimator resulting from the best linear predictor in the scattering domain will not satisfy \eqref{eq:measurements}. More precisely, since the involved distributions are not Gaussian and $\Phi$ is not invertible, the best linear estimator (LMMSE)\footnote{Although it is common to call this estimator linear, it is in fact affine.} will in general not satisfy measurements \eqref{eq:measurements} even with $b = 0$.  That is, letting $\affest$ be the estimator, the set
\[
    {\cal L} \bydef \set{z \ : \ \Phi z = \affest \Phi \Gamma^+ y \wedge \Gamma z = y}
\]
could be empty, and in principle it will. To resolve this issue, we define the estimator $\wh{x}(y)$ by the following three relations (strictly speaking, $\wh{x}$ is a set):
\begin{align}
    \label{eq:cond1_intr}
    &\norm{\Gamma \wh{x} - y } \leq \epsilon, \tag{A1} \\
    \label{eq:cond2_intr}
    &\E [\Phi \wh{x}] = \E [\Phi x], \tag{A2} \\
    \label{eq:cond3_intr}
    &\E [(\Phi \wh{x})_i (\Phi \wh{x} - \Phi x)] = \vec{0}, \ \forall i. \tag{A3}
\end{align}

Let us explain these conditions. The first one is a natural requirement that our estimate satisfy the measurement constraint (\ref{eq:measurements}) up to the noise ball, where $\epsilon^2$ is proportional to variance of the noise $b$. For convenience, we denote by $\cal A$ the set of vectors satisfying the measurement constraint,

\begin{equation*}
    {\cal A} \bydef \set{u \ : \ \norm{\Gamma u - y} \leq \epsilon}.
\end{equation*}

The second and the third condition are related to linearity of estimation. This is not to say that the defined $\Phi \wh{x}$ is linear in $\Phi \Gamma^+ y$---it is somewhat more subtle: we ask that our $\Phi$-domain estimate $\Phi \wh{x}$ optimally exploits the statistical information about $x$ in the sense that no linear (affine) estimator $Q$ applied to $\Phi \wh{x}$ can extract additional information about $\Phi x$ that would make $\E \norm{\Phi x - Q \Phi \wh{x}}^2$ smaller than $\E \norm{\Phi x - \Phi \wh{x}}^2$. This is equivalent to asking that the error vector $\Phi(\wh{x}) - \Phi(x)$ have zero mean and be uncorrelated with every coordinate of the estimate $\Phi \wh{x}$.

The goal is thus to find an estimator $\wh{x}$ which satisfies conditions (\ref{eq:cond1_intr},\ref{eq:cond2_intr},\ref{eq:cond3_intr}). In the next section we propose a new algorithm to compute such an estimator.



\section{Algorithm} 
\label{sec:algorithm}

\subsection{Projection on the Transform-Domain Statistics} 
\label{sub:projection_on_phi_stat}

As argued above, the LMMSE estimator of $\Phi x$ given $\Phi \Gamma^+ y$ will in general not be consistent with the measurements. This means that the relations (\ref{eq:cond1_intr},\ref{eq:cond2_intr},\ref{eq:cond3_intr}) do not specify the LMMSE estimator. However, we propose to use the LMMSE estimator as a building block of an iterative algorithm that will lead to $\wh{x}$ satisfying the three conditions.

We require our estimate to be uncorrelated with the estimation error in the $\Phi$-domain. Thus, given some $\Phi$-domain data $Z = f(y)$ which we interpret as an intermediate estimate (say $Z = \Phi \Gamma^+ y$), we want to produce a random vector $\wh{X} = \wh{X}(Z)$ such that
\begin{align}
    &\E [\wh{X}] = \E [X] \label{eq:centered_trans} \\
    \text{and} \quad &\E [\wh{X}_j (\wh{X} - X)] = 0, \ \forall j \in \set{1, \ldots, n}. \label{eq:noncorr_trans}
\end{align}
Note that there is no typo in having $\wh{X}$ in both terms of the orthogonality relation \eqref{eq:noncorr_trans}. This is because our goal is not to specify how $\wh{X}$ is related to this particular $Z$. Rather, we are stating a property of the estimator $\wh{X}$. Showing how to obtain from $Z$ a vector $\wh{X}$ is then only a mechanism to satisfy this property which we refer to as projection.


Regardless of whether or not $Z$ and $X$ are jointly Gaussian, a random vector satisfying the above conditions can always be found by linear regression over coordinates of $Z$ which optimizes the MSE. In other words, if $\mG \in \R^{n \times n}$, $\vh \in \R^n$ are (uniquely) defined as

\begin{equation}
    \mG, \, \vh = \argmin_{\wh{\mG},\ \wh{\vh}} \ \E \ \norm{X - (\wh{\mG} Z + \wh{\vh})}^2,
\end{equation}
then $\mG Z + \vh$ satisfies \eqref{eq:centered_trans} and \eqref{eq:noncorr_trans}. Conveniently, $\mG$ and $\vh$ only depend on the first and second moments of $(X, Z)$ and we can find them in closed form:

\begin{equation}
    \mG = \mK_{XZ} \mK_{ZZ}^{-1}, \quad \vh = \E [X] - \mG \E [Z].
\end{equation}

\subsection{Projection on the Measurements} 
\label{sub:projection_on_the_measurements}


Given $z \in \R^{d}$ we define a second projector $z' = \proj_{\cal A} z$ where $\Phi z'$ is such that $z' \in {\cal A}$ and $\E \,
\norm{\Phi z' - \Phi z}^2$ is minimized. We can obtain such a projector on a per-realization basis as follows,


\begin{equation}
    \label{eq:measproj_sig}
    z' \in \proj_{\cal A} z \bydef \argmin_{u \in {\cal A}}~\norm{\Phi u - \Phi z}^2.
\end{equation}
Since $\Phi$ is non-linear this projection is non-convex. We can associate to $\proj_{\cal A}$ a projector in the $\Phi$-domain. Given $Z = \Phi z$, we
have

\begin{equation}
    Z' \in \proj_{\Phi(\cal A)} Z \bydef \set{ \Phi v \ : \ v \in \argmin_{u \in {\cal A}}~\norm{\Phi u - Z}^2}.
\end{equation}


\vspace{-2mm}
\subsection{Algorithm} 
\label{sub:algorithm}

We now state the alternating projection algorithm. Let $z^{(0)} = 0$. Then for any
$k \geq 0$ we define
\[
    \Phi z^{(k+1)} \in \proj_{\Phi(\cal A)} \affest \, \Phi z^{(k)} \, .
\]
At each iteration, the linear regression operator $\affest$ applied to $\Phi z^{(k)}$ is computing the closest $Z^{(k+1)}$ such that $\Phi x - Z^{(k+1)}$ is uncorrelated with $\Phi z^{(k)}$. Note that to compute $\proj_{\Phi(\cal A)}$, we must first compute $\proj_{\cal A}$, so the algorithm also maintains an estimate in the signal domain, $z^{(k)}$. The operator $\proj_{\Phi(\cal A)}$ is a non-linear operator which is approximated with a projected gradient descent algorithm. The alternating projection can be explicitly written as
\[
    \Phi z^{(k+1)} \in \proj_{\Phi(\cal A)} \left[ \mG^{(k)} \Phi z^{(k)} + \vh^{(k)} \right].
\]
The algorithm is stopped after $r$ iterations and the associated (non-unique) $z^{(k)}$ is taken as an approximation of $\wh{x}$. It is important to point out that since in every iteration we compute a new estimator, we need not only the first- and second-order statistics of $(\Phi x, \Phi \Gamma^+ y)$, but also of $(\Phi x, \Phi z^{(k)})$ for all iterations $k$.


\section{Numerical Results} 
\label{sec:numerical_results}

\vspace{-1mm}
\subsection{Figures of Merit} 
\label{sub:figures_of_merit}

The proposed algorihtm iteratively reduces the MSE in the $\Phi$ domain. However, this does not translate into improved PSNR in the pixel domain. In fact, in super-resolution, a simple low-pass projection will generally yield better PSNR than our reconstructions. One of the reasons for this is that we reconstruct inexact phase at high frequencies. This results in a slight misalignement of singularities such as points and edges, which in turn leads to large $\ell^2$ errors.  Lowpassing hedges its bets by blurring singularities thus reducing the $\ell^2$ error, although the final result cleary \emph{looks} wrong.

There is a need for a different translation-invariant measure which which goes beyond Gaussian statistics, and which is independent of our specific $\Phi$. To this end, we propose to use higher-order moments, and in particular kurtosis.

\begin{figure*}[t!]
\centering
\includegraphics[width=.95\linewidth]{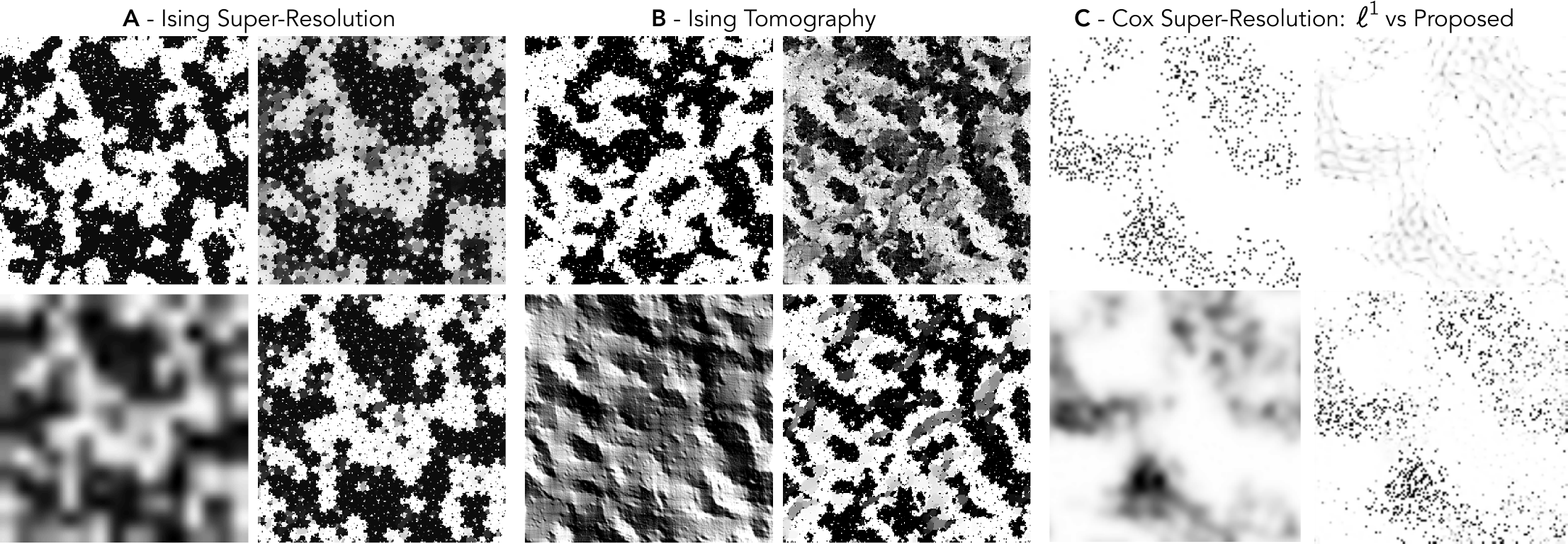}
\caption{Results of computer experiments. \textbf{A}: upper-left---the original Ising realization of size $256 \times 256$, at temperature $T=0.3$; lower-left---lowpass version (projection of the original on $\cal A$) obtained by decimating by a factor of 16 along each axis; upper-right---reconstruction after the first iteration; lower-right---reconstruction after 20 iterations, projected on $\cal A$. \textbf{B}: same as \textbf{A} but with half-angle Radon measurements. \textbf{C}: upper-left---realization of a Cox random process (non-homogeneous Poisson); lower-left---lowpass version; upper-right---reconstruction by $\ell^1$ minimization; lower-right---one iteration of the proposed algorithm.}\vspace{-2mm}
\label{fig:results}
\end{figure*}


Kurtosis of a random variable is the standardized fourth central moment. For a random vector $\vx = [x_1, \ldots, x_N]$, one may define a multivariate (co)kurtosis tensor by analogy to the covariance matrix:

\begin{equation}
    \kurt[x]_{ijk\ell} = \frac{\E[(x_i - \mu_i)(x_j - \mu_j)(x_k - \mu_k)(x_\ell - \mu_\ell)]}{\sqrt{\var[x_i]\var[x_j]\var[x_k]\var[x_\ell]}}.
\end{equation}
Note that for shift-invariant processes we can fix one index.

To circumvent the challenges of estimating and comparing large tensors, we further use a result by Mardia \cite{Mardia:1970by} who showed that $\beta \bydef \E\left\{ \left[(x - \mu)^\T \mSigma^{-1} (x - \mu)\right]^2 \right\}$ is a good measure of multivariate kurtosis. Given a random sample $x_1, \ldots, x_n$, we can estimate $\beta$ by the empirical mean using estimated covariance $\wh{\mSigma}$. Mardia demonstrated that for $x \in \R^p$ we have $\E[\wh{\beta}] = p(p+2) \frac{n - 1}{n + 1}$.

Non-normality is an important aspect of natural signals. For an illustration in texture classification see \cite{Bruna:2013hy}. We will use the excess sample kurtosis $\wh{\beta} - p(p+2)$ as a measure of how well the statistics of the desired signals are reproduced by the various methods.






\vspace{-2mm}
\subsection{Operators and Processes} 
\label{sub:operators}

We study two different problems: super-resolution and limited-angle tomography (Radon transform). Their action in Fourier space is illustrated in Fig. \ref{fig:known_spectrum}. In super-resolution, the operator is given as a decimator---a composition of lowpass filtering $H_{\mathrm{LP}}$ and downsampling $S$, $\Gamma_\mathrm{SR} = S \circ H_{\mathrm{LP}}$,
and the corresponding projection is given as (assuming $y$ are measurements and $H_\mathrm{LP}$ is self-adjoint)
\begin{equation}
    P_\mathrm{SR} z = H_\mathrm{LP} \circ S^* y + (I - H_\mathrm{LP}) z,
\end{equation}
where $S^*$ is upsampling.

In tomography, the operator is given as the Radon transform $[\Gamma_\mathrm{R} x]_\ell = \int_\ell x(s) \, \di s,$
for a set of lines $\ell \in \Lambda$. Unlike in the case of super-resolution, here we do not use a linear right inverse. A better result can be obtained if we impose positivity and max norm:

\begin{equation}
    \label{eq:radon_proj}
    \Gamma_{\mathrm{R}}^+ y \bydef \argmin_{z : 0 \leq z(u) \leq 1} \norm{y - \Gamma_\mathrm{R} z}_2^2,
\end{equation}
Note that this is a right inverse only on signals in the constraint space.

We use two types of signals: realizations of the Ising spin-glass model \cite{Binder:1986ej} and Cox point processes \cite{Cox:1980vt}. At the right temperature, Ising realizations behave like a mixture of simple shapes and textures while Cox processes are useful for comparisons with sparse regularizations.


\vspace{-2mm}

\subsection{Comparison With $\ell^1$ Minimization: Cox Point Processes} 
\label{sub:comparison_with_l1}

We first attempt to demonstate why convex methods that work in the pixel domain and do not use a phase-removal mechanism fail in some simple cases. Fig. \ref{fig:results}\textbf{C} shows a reconstruction of a Cox point process. Measurements are obtained by Gaussian filtering and downsampling by a factor of 4 along each axis. Signal sparisty is set so that combined with the relatively high coherence of the forward operator the $\ell^1$ minimization does not have a unique solution \cite{Donoho:2006ci}.

Convexity of the cost function $E(u) \bydef d( y, \Gamma u) + \lambda h(u)$ can be a bottleneck in severely ill-posed problems as it makes the estimate vulnerable to regression to the mean. If $u_1$ and $u_2$ are two minimizers with $E(u_1) = E(u_2)$, then $\tilde{u} = t u_1 + (1-t) u_2$ is also a minimizer of $E$. But if $u_2$ is a small translation or deformation of $u_1$,  $\tilde{u}$ will be a blurred version of $u_1$ with corrupted high-frequency information. This can be seen clearly for $h(\, \cdot \,) = \norm{\, \cdot \,}_{\ell^1}$ in Fig. \ref{fig:results}\textbf{C}.

The reconstruction using one iteration of the proposed algorithm looks better. Even though it is not equal to the original realization, it has the correct spatial statistics (it \emph{looks} right), and it reproduces measurements. It is thus a valid solution within the proposed framework.
Indeed, Table \ref{tbl:fom} shows that although the $\ell^1$-minimal solution fares better in terms of the MSE, in terms of higher-order moments it performs much worse than the scattering-based reconstruction.



\begin{figure}[t]
\centering
\includegraphics[width=.70\linewidth]{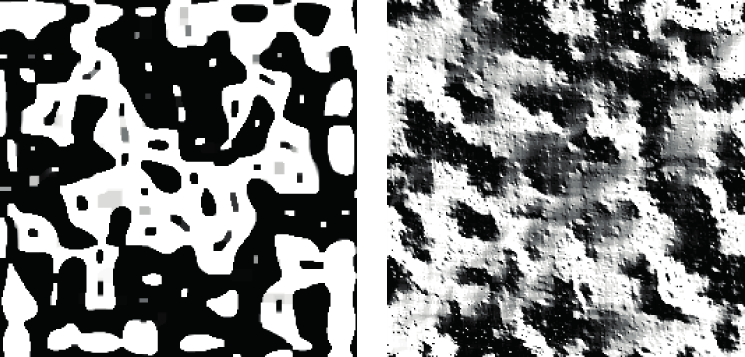}
\caption{Reconstructions of Ising measurements from Fig. \ref{fig:results} by TV-norm minimization. Left: super-resolution; right: tomography.}
\label{fig:tvrec}
\end{figure}

\begin{figure}[t]
\centering
\includegraphics[width=0.9\linewidth]{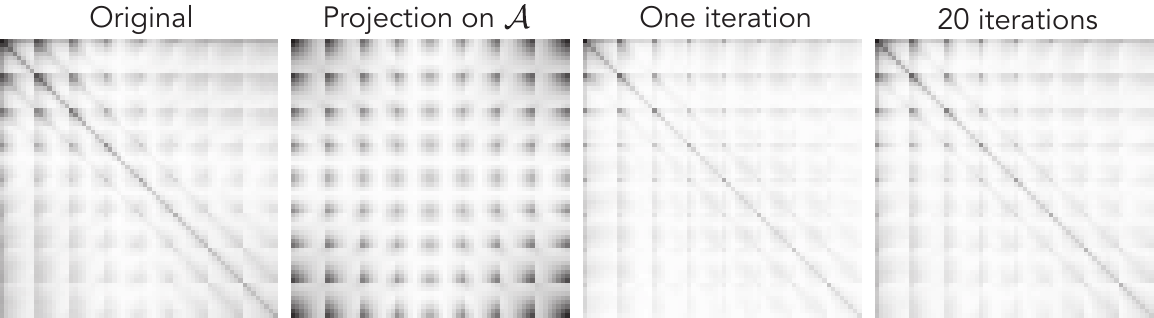}
\caption{Translation invariant cokurtosis tensors of size $64 \times 64$ computed on patches of size $8 \times 8$ and averaged along the first dimension (tomography).}
\label{fig:cokurtosis}
\vspace{-2mm}
\end{figure}

\begin{table}[t]
\centering
\scalebox{.85}{
\begin{tabular}{@{} l l l @{} }
    \toprule
    \textbf{Experiment} & \textbf{MSE} & \textbf{Ex. Kurtosis}  \\ 
    \midrule
    Ising Original          & 0       & 1760  \\
    \vspace{-1mm}\\
    Ising SR Projection     & 0.91e4  & 85    \\
    Ising SR Iteration 1    & 1.32e4  & 1282  \\
    Ising SR Final          & 1.35e4  & 1290  \\
    \vspace{-1mm}\\
    Ising Radon Projection  & 6.91e3  & 2285  \\
    Ising Radon Iteration 1 & 6.55e3  & 1705  \\
    Ising Radon Final       & 7.54e3  & 1787  \\
    \vspace{-1mm}\\
    Cox Original            & 0    & 605      \\
    Cox SR Projection       & 1814 & 267      \\
    Cox $\ell^1$            & 1500 & 3164     \\
    Cox One Iteration       & 1910 & 514      \\
    \bottomrule
\end{tabular}}
\caption{Figures of merit computed on various sets of images.}
\label{tbl:fom}
\vspace{-1mm}
\end{table}

\vspace{-2mm}
\subsection{Ising Super-Resolution} 
\label{sub:ising_superresolution}

Fig. \ref{fig:results}\textbf{A} shows super-resolution of Ising realizations. The operator is a decimation by a factor of 16 along each axis, thus the data loss is 256-fold. The following can be noted: using a linear reconstruction (lower-left) gives a bad result. Already the first iteration of the proposed algorithm gives a much better result, and iterating further brings out correct structural details as indicated in the lower right. In Table \ref{tbl:fom} we see that the excess kurtosis corroborates our ocular observations, although the MSE might suggest otherwise. In the light of the severity of data loss we find these results remarkable.

For a comparison, in Fig. \ref{fig:tvrec} we provide a reconstruction regularized by the total variation (TV):
$\argmin_{z : 0 \leq z(u) \leq 1} \norm{y - \Gamma_\mathrm{SR} z}_2^2 + \lambda \normtv{z}.$
The value $\lambda = 5 \times 10^{-6}$ was tuned by hand.


\vspace{-1mm}

\subsection{Ising Tomography} 
\label{sub:ising_tomography}

In Fig. \ref{fig:results}\textbf{B} we present reconstructions of Ising models from tomographic (Radon) measurements. $\Gamma_\mathrm{R}$ performs a half angle Radon transform with angles between $0^\circ$ and $89^\circ$ in steps of $1^\circ$. We see that a simple reconstruction subject to the positivity constraint \eqref{eq:radon_proj} does not give a very good result. First iteration of our algorithm gives a nicer image, and a considerable improvement comes from iterating.

The excess kurtosis given in Table \ref{tbl:fom} corroborates what our eyes are telling us. We also show the cokurtosis tensors in Fig. \ref{fig:cokurtosis}. It is clear that the final result best matches the cokurtosis of the original signal.

Finally, a TV-norm regularized reconstruction is shown in Fig. \ref{fig:tvrec} ($\lambda = 1\times 10^{-6}$). Since in the tomographic experiments the data loss is less severe, this reconstruction looks better than in super-resolution.



\vspace{-2mm}

\section{Conclusion} 
\label{sec:conclusion}

We proposed a new algorithm for solving linear inverse problems which exploits the correlation structure of the data in a non-linearly transformed domain. The modulus non-linearity reveals spectral correlations otherwise hidden by phase fluctuations. Promising initial results on super-resolution and tomography show that we indeed recover the correct missing spectral information in hard problems. Future work includes optimizing the transform, more general classes of operators, efficient implementations, and proofs of convergence.


\bibliographystyle{IEEEbib}
\bibliography{scatinv}

\end{document}